\documentclass[11pt,twoside]{article}

\usepackage{ihepconf}

\def\be{\begin{equation}}
\def\ee{\end{equation}}

\input{amssym.def}
\input{amssym}

\input{amssym.def}
\input{amssym}

\begin{document}

  \title{\LARGE \bf Einstein, Hilbert and Equations of Gravitation}
 \vskip 5mm
 \author{{\large\bf V.A. Petrov}\\ [1mm]
 {\it Institute for High Energy Physics, 142200, Protvino,
 Russia}}

  \maketitle
 
$\bullet$ The history of the equations of gravitational field is of
particular interest because they are related to two famous names:
Albert
Einstein and David Hilbert, and because the circumstances of their
invention bear sometimes almost detective character. 

One of the early references is the book ``Einstein, Hilbert and the
Theory of Gravitation'' by a renowned historian of science
J.~Mehra~[1], where the great role of Hilbert was showed very
clear. Such view was strengthened in 1978 when the correspondence
between Einstein and Hilbert was published, from which followed that
Hilbert informed Einstein on the gravitational field equations in a
letter before his formal publication~[2].

The opinion, shared by many physicists, was that it was Hilbert who
possessed an undoubtful priority~[3]. 

$\bullet$ However, in 1997 a new sensation shaked just established
opinion: the authors of a short article in ``Science''~[4] argued on
the basis of the first proofs of the Hilbert paper on the
gravitational equations, digged up from the Hilbert archive, 
that {\bf Hilbert had no correct, generally covariant equation before
Einstein.} Moreover, the authors of~[4] transparently alluded that
Hilbert ``borrowed'' some decisive formulae from Einstein! And even
that Hilbert tried to hide such an appropriation with help of
deliberately wrong dating of his article.

Such an accusation would seriously undermine the image of David
Hilbert from the ethical side, and was in a sharp contrast to all
what was known about his personality.

$\bullet$ On the other hand the very personality of Einstein is by no
means irreproachable. Take, for instance, the case of the relativity
theory. So, counter-reaction to paper~[4] followed. 

One of the first
was the book by C.J.~Bjerknes~[5], well documented and with a rich
bibliography, in which the conclusions of paper~[4] were contested.
This
was based on the correspondence between Einstein and Hilbert and an
important fact that the proofs of Hilbert's paper --- the main
evidence of the authors of~[4] against Hilbert --- were mutilated
with some part of proofs cutted off.

This fact was mentioned for the first time, but without a due
evaluation, in Ref.~[6]. In book~[5] it was mentioned that
F.~Winterberg assumed that the explicit form of the gravitational
field could be fairly contained in the cutted off parts of the
proofs. This seriously undermined the main argument of the
authors~[4] against Hilbert\footnote{Afterwards this consideration
was published~[7], and, finally, the book~[8] appeared in which
D.~Wuensch gave a thorough analysis of the mutilated proofs and
other relevant documents with a conclusion: Hilbert knew the explicit
form of the gravitational equations, they contended in the proofs 
and the latter were deliberately mutilated in order to falsify the
historical truth. --- {\it Note added to proof.}}.

$\bullet$ In ref.~[9] the question was considered in detail with
analysis of the Hilbert and Einstein papers. 

Here I will reproduce only one simple reason we used to reject one of
the main accusation of the authors~[4] against Hilbert, namely:
{\it ``...knowledge of Einstein's result may have been crucial to
Hilbert's introduction of the trace term into his field equations''.
}

Thus, according to the authors of~[4], Hilbert's  equations
$$
\frac{\delta \sqrt{g}H}{\delta g^{\mu\nu}}=0,\eqno{(*)}
$$
where $H$ is the sum of the gravitational and material lagrangians,
$g^{\mu\nu}$ is the metric tensor with the determinant  $-g$, are
incomplete and  need  some ad hoc ``introduction'' of
additional terms (``trace term''). This ``discovery'' of the authors
of
paper~[4] clearly demonstrates their professional inconsistency:
they
must be never tried to calculate the variational derivative (*)
themselves! Otherwise they would quickly saw that the ``crucial''
trace
term is safely contained in Eq.(*).

Other mistakes of the authors of paper~[4] where analysed in our paper~[9]. 

One thing has to be clear. {\bf At the very moment when Hilbert
equated
the gravitational part of the lagrangean to the  scalar Riemannian
curvature the whole game was over. }

All the rest was the matter of almost routine calculations, though
one has to mention that Hilbert managed to get, in his paper~[10] 
some very important
results (``Bianchi identity''). All detailes are given in~[9]. 

So, we conclude that an unfair attack of the authors of ref.~[4]
against Hilbert's originality in deriving the gravitaional field
equations is completely and shamefully failed.

$\bullet$ Now, what about Einstein? In ref.~[9], there was
admitted that  Einstein could derive the gravitational field
equations~[11] independently of Hilbert. 

The main evidence, if to be as loyal as possible to Einstein, is his
assertion, made by him in the letter to Hilbert of 18~November 1915:
{\it ``The system you furnish agrees --- as far as I~can --- exactly
with what I~found in the last few weeks and presented to the
Academy''}~[2].

That is, Einstein, in this letter, acknowledged receiving the
Hilbert equations of the gravitational field and informed him that
his, Einstein's, equations are essentially the same. 

But we know all papers by Einstein presented to the Academy ``in the
last few weeks'',
including the paper of the 18~November --- they all are still wrong
and do not contain the trace term. One has to concede that Einstein
had by 18~November 1915 the correct equations but preferred to
publish the wrong ones up to 25~November! In principle this is
possible, but...

We also have to mention the  paper by Einstein of
18~November 1915~[12] where he claimed the successful test of his
(wrong!)
theory in obtaining the correct result for the Mercury perihelion.
F.~Winterberg (as cited in ref.~[5]) draw the attention to the fact
that if
Einstein would really follow his equations as they were described in
his paper of 18~November, he would obtain the result twice larger
than the correct one. Nonetheless, his final result was correct!
Further interesting details of this story can be found in~[5].

$\bullet$ What is our conclusion? We still keep the opinion expressed
in~[9]: the gravitational equation has to be named as
the ``Einstein--Hilbert equation''. The reason is that it was
Einstein
who posed the problem to find out the equation in which the
energy-momentum tensor is a source for gravitational potentials~[13].
Hilbert had found such an equation. Einstein derived it, quite
probably, later in
his
own way.

\end{document}